\pdfoutput=1
\documentclass[10pt,conference]{IEEEtran} 
\IEEEoverridecommandlockouts 
\usepackage{cite}
\usepackage{amsmath,amssymb,amsfonts}
\usepackage{algorithmic}
\usepackage{graphicx}
\usepackage{balance}
\usepackage{textcomp}
\usepackage{wrapfig}
\usepackage[switch]{lineno}

\usepackage[table,xcdraw]{xcolor}
\usepackage{multirow}
\usepackage{subcaption}
\usepackage{tcolorbox}
\usepackage{dblfloatfix}
\usepackage{listings}

\definecolor{codegreen}{rgb}{0,0.6,0}
\definecolor{codegray}{rgb}{0.5,0.5,0.5}
\definecolor{codepurple}{rgb}{0.58,0,0.82}
\definecolor{backcolour}{rgb}{0.95,0.95,0.92}

\lstdefinestyle{mystyle}{
  backgroundcolor=\color{backcolour},   commentstyle=\color{codegreen},
  keywordstyle=\color{magenta},
  numberstyle=\tiny\color{codegray},
  stringstyle=\color{codepurple},
  basicstyle=\ttfamily\footnotesize,
  breakatwhitespace=false,         
  breaklines=true,                 
  captionpos=b,                    
  keepspaces=true,                 
  numbers=left,                    
  numbersep=5pt,                  
  showspaces=false,                
  showstringspaces=false,
  showtabs=false,                  
  tabsize=2
}

\usepackage{url}

\newcommand{\hil}{\cellcolor[gray]{.9}}

\newcommand{\rc}{\rowcolor[gray]{.9}}

\newcommand{\all}{\cellcolor[HTML]{85C1E9}}
\newcommand{\rr}{\cellcolor[HTML]{F4D03F}}

\newcommand{\early}{\cellcolor[HTML]{ABEBC6}}

\usepackage[tikz]{bclogo}
{\noindent\begin{minipage}[c]{\linewidth}%
\begin{bclogo}[couleur=gray!10,%
    arrondi=0.1,%
    logo=\bctrombone,%
    ombre=true]{{\small ~#1}}}%
{\end{bclogo}\end{minipage}}

\definecolor{recency}{HTML}{F4D03F}
\definecolor{all}{HTML}{85C1E9}
\definecolor{early}{HTML}{ABEBC6}

\definecolor{lower}{HTML}{FF9933}
\definecolor{higher}{HTML}{82E0AA}

\newcommand{\tion}[1]{\S\ref{tion:#1}}
\newcommand{\fig}[1]{Figure~\ref{fig:#1}}
\newcommand{\tbl}[1]{Table~\ref{tbl:#1}}

\newcommand{\bi}{\begin{itemize}}
\newcommand{\ei}{\end{itemize}}

\def\BibTeX{{\rm B\kern-.05em{\sc i\kern-.025em b}\kern-.08em
    T\kern-.1667em\lower.7ex\hbox{E}\kern-.125emX}}
\begin{document}

\title{Early Life Cycle Software Defect Prediction.\\Why? How?}

\author{
 \IEEEauthorblockN{N.C. Shrikanth, Suvodeep Majumder and Tim Menzies}
 \IEEEauthorblockA{Department of Computer Science, North Carolina State University, Raleigh, USA\\ snaraya7@ncsu.edu, smajumd3@ncsu.edu, timm@ieee.org}
}

\maketitle

\begin{abstract}
 Many researchers assume that, for software analytics,  ``more data is better.'' We write to show that, at least for learning defect predictors, this may not be true.
 
 To demonstrate this,  we analyzed hundreds of popular GitHub projects. These projects ran for 84  months and contained   3,728  commits (median values).
Across these projects, most of the defects occur very early
in their life cycle.
Hence,
defect predictors learned from the first
150 commits and four months  perform
just as well as anything else.
This means that, at least for the projects studied here,  after the first few months, we need not continually update our defect prediction models.

We hope these results inspire other researchers to adopt a ``simplicity-first'' 
approach to their work. 
Some domains require
a complex and data-hungry analysis. 
But before   assuming
complexity, it is prudent to check   the raw data looking for 
``short cuts'' that can simplify the   analysis.

\end{abstract}

\begin{IEEEkeywords}
sampling, early, defect prediction,  analytics
\end{IEEEkeywords}

\section{Introduction} \label{tion:introduction} 

This paper proposes a  {\em data-lite} method that finds effective software defect predictors using data just from the first 4\% of a project's lifetime. 
Our new method is recommended since it means that we need not always revise defect prediction models, even if new data arrives.
This is important since
 managers, educators, vendors, and researchers lose faith in methods that are always changing their conclusions.

\begin{figure*}[!t]
\includegraphics[width=\linewidth]{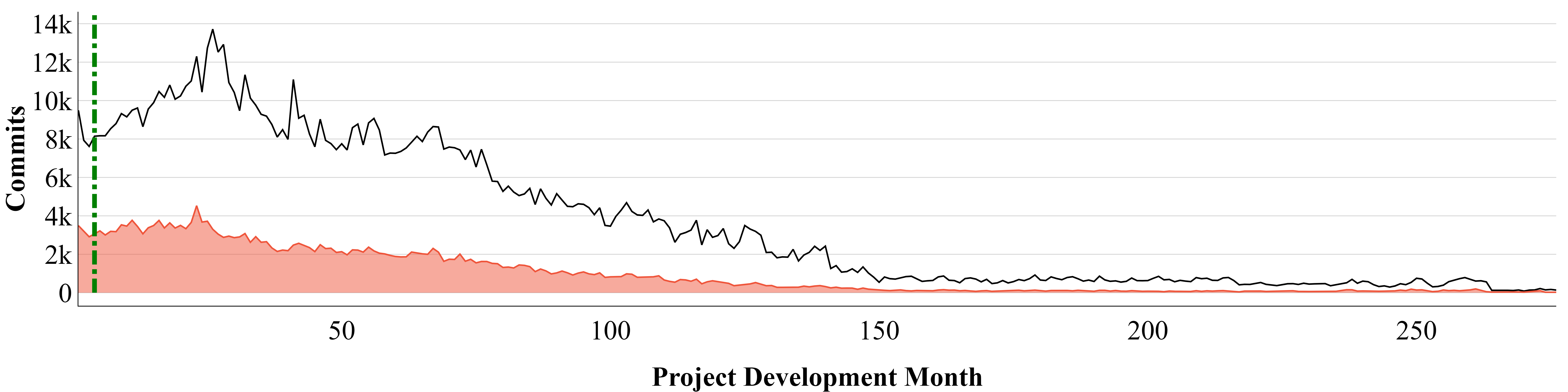}
\caption{1.2 million commits for 155 GitHub projects. Black:Red (shaded) = Clean:Defective commits. In this paper, we compare (a) models learned up to the vertical green (dotted) line to (b)   models learned using more data.}\label{fig:whale}
\end{figure*}
Our method is somewhat unusual since it takes
an opposite approach to {\em data-hungry}
methods that  (e.g.)   
 use   data collected across  many  years of a software project~\cite{d2012evaluating,kamei2012large}.  
Such  data-hungry methods are often cited as
the key to success for data mining applications.
For example,
in his famous talk, ``The Unreasonable Effectiveness of Data,'' Google's Chief  Scientist Peter Norvig
argues that 
``billions of trivial data points can lead to understanding''~\cite{norvig11}
(a claim he supports with  numerous
examples from vision research).

But what if some Software Engineering (SE) data was not like Norvig's data? What if SE
needs its own AI methods, based on what we learned about
the specifics of software projects? If that were true, then data-hungry methods might be needless over-elaborations of a fundamentally simpler process.

This paper shows that for one specific software analytics task (learning defect predictors), we do not need a data-hungry approach.
We observe in \fig{whale} that while the median lifetime of many projects is 84 months, most of the defects from those projects occur much earlier than that. That is, very little of the defect experience occurs later in the life cycle.
 Hence, predictors learned after 4 months (the vertical green dotted  line in \fig{whale}) do just as well as anything else;
i.e.   learning can stop after just    $4$\% of the life cycle (i.e., 4/84 months).
That is to say, when learning  defect predictors:

\begin{center}
{\em
96\% of the time,  we do not  want and\\
we do not  need data-hungry methods}.
\end{center}
We stress that we have only shown an ``early data is enough'' effect
in the special case of
(a)~defect prediction  for
(b)~long-running non-trivial engineering GitHub projects studied here (what Munaiah et al.\cite{munaiah2017curating} would call ``organizational projects'').
Such projects can be readily identified by how many ``stars'' (approval marks) they have accumulated from GitHub users.
Like other researchers (e.g., see the 
TSE'20 article by Yan et al.~\cite{yan2020just}),
we explore projects with at least 1000 stars.

That said, 
 even  within these restrictions,  we believe we are exploring an
interesting range of projects. 
Our sample includes numerous widely used applications developed by Elastic (search-engine\footnote{https://github.com/elastic/elasticsearch}), Google (core libraries\footnote{https://github.com/google/guava}), Numpy (Scientific computing\footnote{https://github.com/numpy/numpy}),  etc. Also, our sample of projects is written in widely used programming languages, including C, C++, Java, C\#, Ruby, Python, JavaScript, and PHP.

Nevertheless, 
in future work, we need
to explore the external validity of our results to other SE tasks (other than defect prediction) and for other kinds of data.  For example, 
Abdalkareem et al.~\cite{Abdalkareem2020} show that up to 16\% of Python and JavaScript packages are ``trivially small'' (their terminology); i.e., have less than 250 lines of code. It is an open issue if our methods work for
other kinds of software such as those trivial Javascript and Python packages. To support such further explorations, we have placed all our data, scripts   on-line\footnote{ For a replication package see: \url{https://doi.org/10.5281/zenodo.4459561}}.

The rest of this paper is structured as follows. In \S2, we discuss the negative
consequences of excessive data collection, then in \S3 we show that for hundreds of GitHub~\cite{cite_github} projects, the defect data from the
latter life cycle defect data is relatively uninformative.
This   leads to the definition of  experiments in the early life cycle
defect prediction  (see \S4,\S5). 
From those experiments (in \S6), we show that at least for defect prediction,   a small sample of data is useful, but (in contrast to Norvig's claim)  more data is 
\underline{{\em not}} more useful. Lastly, \S7 discusses some threats, and conclusions are presented in \S8.

\section{Background}

 \subsection{About Defect Prediction}
 
 Defect prediction
uses  data miners to input static code attributes and 
output models that  predict where
  the   code  probably contains   most bugs~\cite{ostrand2005predicting,menzies2006data}.
Wan et al.~\cite{wan2018perceptions}  reports   that there   is much industrial interest in these predictors since they can guide the deployment of more  expensive and  time-consuming quality assurance methods (e.g., human inspection). 
Misirili et al~\cite{misirli2011ai} and Kim et al.~\cite{kim2011empirical} report
considerable cost savings when such predictors are used in guiding industrial quality assurance processes. 
Also, Rahman et al.~\cite{10.1145/2568225.2568269} show that such predictors are competitive with more elaborate approaches.

  In defect prediction,
 data-hungry researchers assume that if data is useful, then even more data is much more useful. For example:  
 \bi
 \item
  ``..as long as it is large; the resulting prediction performance is likely to be boosted more by the size of the sample than it is hindered by any bias polarity that may exist''~\cite{rahman2013sample}.
  \item
``It is natural
to think that a closer previous release has more similar characteristics and
thus can help to train a more accurate defect prediction model. It is also
natural to think that accumulating multiple releases can be beneficial because
it represents the variability of a project''~\cite{amasaki2020cross}. 
\item
 ``Long-term JIT models should be trained using a cache of
plenty of changes''~\cite{mcintosh2017fix}.
\ei

Not only are   researchers hungry for data,
but they are also most hungry for the most recent data.
For example:
Hoang et al. say 
 ``We assume that older commits changes may have characteristics that no longer effects to the latest commits''~\cite{8816772}.
Also,
it is common practice in defect prediction to perform ``recent  validation'' where predictors are tested on  the latest release after training from the prior one or two
 releases~\cite{tan2015online,mcintosh2017fix,kondo2020impact,fu2016tuning}. 
For a project with multiple releases,  recent validation ignores any insights that are available from older releases.

 \subsection{Problems with Defect Prediction: ``Conclusion Instability''}\label{problems}
 
If we  revise old models whenever new data becomes available, then this can lead to  ``conclusion instability'' (where new data leads to different models).  
 Conclusion instability is well documented. Zimmermann et al.~\cite{zimmermann2009cross} learned defect predictors from 622 pairs of projects (project1, project2). In only 4\% of pairs, predictors from project1 worked on project2. Also, Menzies et al.~\cite{Me13} studied defect prediction results from 28 recent studies, most of which offered widely differing conclusions about what most influences software defects. Menzies et al.~\cite{menzies2011local} reported experiments where data for software projects are clustered, and data mining is applied to each cluster. They report that very different models are learned from different parts of the data, even from the same projects.
 
In our own past work, we have found conclusion instability, meaning there we had to throw years of data. In one sample of GitHub data, we sought to learn everything we could from 700,000+    commits. The web slurping required for that process took nearly 500  days of CPU (using five machines with 16 cores, over 7 days).
Within that data space, we
found significant differences in the models learned from different parts of the data. So even after all that work,
we were unable to offer our business users a stable predictor
for their domain.

Is that the best we can do? 
Are there general defect prediction principles we can use to guide project management, software standards, education,   tool development, and legislation about software? 
Or is  SE some ``patchwork quilt'' of ideas and methods where it only makes sense to reason about
specific, specialized, and small sets of related projects? 
Note that if the software was a ``patchwork'' of ideas,   then    there would be no stable conclusions about what constitutes best practice for software
engineering (since those best practices would keep changing as we move from project to project). 
Such conclusion instability
would have detrimental
implications for {\em   trust, insight, training}, and {\em tool
          development}.

{\em Trust:}  
Conclusion instability is unsettling for project managers. Hassan~\cite{hassan17} warns that managers lose trust in software analytics if its results keep changing. Such instability prevents project managers from offering clear guidelines on many issues, including (a) when a certain module should be inspected; (b) when modules should be refactored; 
and (c) deciding where to focus on expensive testing procedures.
 
{\em Insight:} 
Sawyer et al. assert that insights are essential to catalyzing business initiative~\cite{sawyer2013bi}. 
From Kim et al.~\cite{Kim2016} perspective, software analytics is a way to obtain fruitful insights that guide practitioners to accomplish software development goals, whereas for Tan et al.~\cite{tan2016defining} such insights are a central goal. From a practitioner's perspective Bird et al.~\cite{Bird:2015} report, insights occur when users respond to software analytics models. Frequent model generation could exhaust users' ability for confident conclusions from new data.
 
{\em Tool development and Training:} 
Shrikanth and Menzies~\cite{shrikanth2019assessing} warns
that unstable models make it hard to onboard novice software engineers. Without knowing what factors  most influence the local project, it is hard to
design and  build appropriate tools for quality assurance  activities

\begin{figure}[!b]
\begin{center}
\includegraphics[width=2.6in]{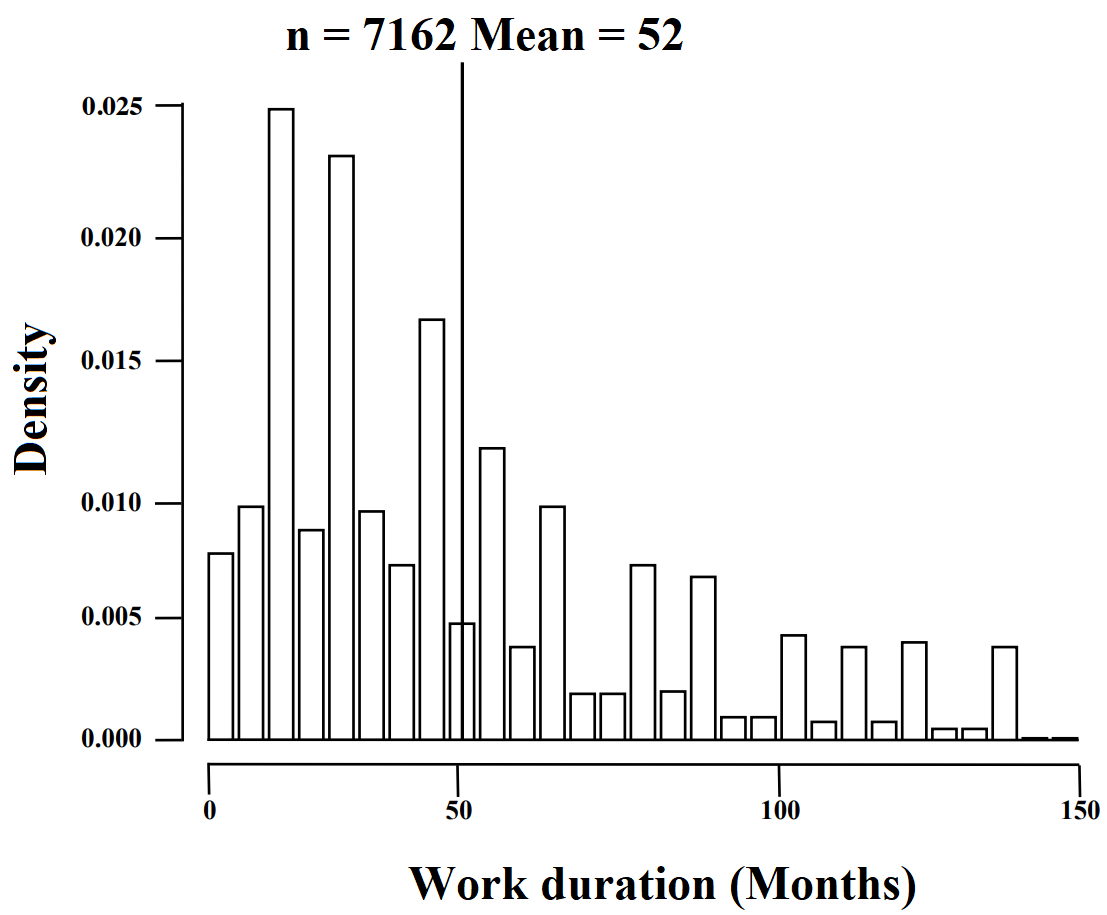}
\end{center}
\caption{Work duration histograms on particular projects;  from~\cite{Sela18}. Data from:  Facebook, eBay, Apple, 3M, Intel and Motorola. }\label{fig:work}  
\end{figure}

All these problems with  trust, insight, training, and tool
          development
          could be solved, if early on in the project, a  defect prediction model can be learned that is effective for the rest of the life cycle.   
 As mentioned in the introduction, we study here GitHub projects spanning 84 months and containing 3,728 commits     (median values). 
 Within that data, 
 we have found that  
 models learned
after just     150 commits (and four months of data collection), perform just as well
as anything else. In terms of resolving conclusion instability, this is a very significant result since
it means that for $4/84=96\%$ of the life cycle, we can offer stable defect predictors.

One way to consider the impact of such early life cycle predictors is to use the data of \fig{work}.
That plot shows that software employees usually change projects every 52 months
(either moving between companies or changing projects within an organization).
This means that in seven years (84 months), the majority of workers and managers would first appear on a job 
{\em after}
the initial four months required to learn a defect predictor.  Hence,  for most workers and managers,
the detectors learned via the methods of this paper would be the ``established wisdom'' and ``the way we do things here'' for their projects.
This means that a detector learned in the first four months would
be a suitable oracle to guide training and hiring;  the development of code review practices; the automation
of local ``bad smell detectors''; as well as tool selection and development.

\begin{table*}[!t]

\scriptsize
 
\begin{tabular}{|lrr|l|r|lrr|l|r|lrr|l|r|}
\hline
\cline{1-15}
\textbf{  \textbf{Paper}} & \textbf{\textbf{Year}} & \textbf{\textbf{Citations}} & \textbf{\textbf{Sampling}} & \textbf{\textbf{Projects}} & \textbf{  \textbf{Paper}} & \textbf{\textbf{Year}} & \textbf{\textbf{Citations}} & \textbf{\textbf{Sampling}} & \textbf{\textbf{Projects}} & \textbf{ \textbf{Paper}} & \textbf{\textbf{Year}} & \textbf{\textbf{Citations}} & \textbf{\textbf{Sampling}}     & \textbf{\textbf{Projects}} \\ \cline{1-15}
~\cite{menzies2008implications}                 & 2008                   & 172                         & \cellcolor[gray]{.9}All    & 12                         & ~\cite{fu2016tuning}                            & 2016                   & 111                         & Release                    & 17                         & ~\cite{agrawal2018better}                       & 2018                   & 61                          & \cellcolor[gray]{.9}All        & 9                          \\
~\cite{zhang2010value}                          & 2010                   & 21                          & \cellcolor[gray]{.9}All    & 5                          & ~\cite{fu2016differential}                      & 2016                   & 28                          & Release                    & 10                         & ~\cite{Rebalancing2018}                                 & 2018                   & 43                          & Percentage                     & 101                        \\
~\cite{menzies2010defect}                       & 2010                   & 347                         & Percentage                 & 10                         & ~\cite{7886923}                                 & 2016                   & 130                         & Release                    & 10                         & ~\cite{8330264}                                 & 2018                   & 15                          & Release                        & 13                         \\
~\cite{6080797}                                 & 2011                   & 94                          & \cellcolor[gray]{.9}All    & 8                          & ~\cite{huang2017supervised}                     & 2017                   & 44                          & \cellcolor[gray]{.9}All    & 6                          & ~\cite{chen2019software}                        & 2019                   & 16                          & Percentage                     & 7                          \\
~\cite{kamei2012large}                          & 2012                   & 264                         & \cellcolor[gray]{.9}All    & 11                         & ~\cite{mcintosh2017fix}                         & 2017                   & 36                          & Month                      & 6                          & ~\cite{pascarella2019fine}                      & 2019                   & 14                          & Month                          & 10                         \\
~\cite{d2012evaluating}                         & 2012                   & 387                         & \cellcolor[gray]{.9}All    & 5                          & ~\cite{nam2017heterogeneous}                    & 2017                   & 220                         & \cellcolor[gray]{.9}All    & 34                         & ~\cite{kondo2019impact}                         & 2019                   & 11                          & Percentage                     & 26                         \\
~\cite{wang2013using}                           & 2013                   & 322                         & \cellcolor[gray]{.9}All    & 10                         & ~\cite{7539677}                                 & 2017                   & 44                          & Percentage                 & 255                        & ~\cite{huang2019revisiting}                     & 2019                   & 14                          & Slice                          & 6                          \\
~\cite{zhang2014towards}                        & 2014                   & 105                         & \cellcolor[gray]{.9}All    & 1,403                      & 
~\cite{OZTURK201717}                            & 2017                   & 8                           & \cellcolor[gray]{.9}All    & 10                         &
~\cite{yan2019characterizing}                   & 2019                   & 1                           & \cellcolor[gray]{.9}All        & 10                         \\
~\cite{6982637}                                 & 2014                   & 44                          & Release                    & 1                          & ~\cite{dam2019lessons}                          & 2018                   & 36                          & \cellcolor[gray]{.9}All    & 11                         & ~\cite{yang2019empirical}                       & 2019                   & 0                           & Percentage                     & 6                          \\
~\cite{10.1145/2568225.2568269}                 & 2014                   & 93                          & Release                    & 5                          & ~\cite{tantithamthavorn2018impact}              & 2018                   & 66                          & Percentage                 &                            & ~\cite{yatish2019mining}                        & 2019                   & 4                           & Release                        & 9                          \\
~\cite{tan2015online}                           & 2015                   & 129                         & \cellcolor[gray]{.9}All    & 7                          & ~\cite{wu2018cross}                             & 2018                   & 33                          & \cellcolor[gray]{.9}All    & 18                         & ~\cite{bennin2019relative}                      & 2019                   & 10                          & Release                        & 20                         \\
~\cite{ryu2016value}                            & 2016                   & 87                          & \cellcolor[gray]{.9}All    & 10                         & ~\cite{wang2018deep}                            & 2018                   & 25                          & \cellcolor[gray]{.9}All    & 16                         & ~\cite{8816772}                                 & 2019                   & 8                           & \cellcolor[gray]{.9}All, Slice & 2                          \\
~\cite{krishna2016too}                          & 2016                   & 42                          & Release                    & 23                         & ~\cite{chen2018multi}                           & 2018                   & 40                          & \cellcolor[gray]{.9}All    & 6                          & ~\cite{kondo2020impact}                         & 2020                   & 0                           & \cellcolor[gray]{.9}All        & 6                \\ \hline

\end{tabular}
 
\caption{Papers discussing different sampling policies. 
All (papers that utilize all historical data to build defect prediction models, shaded in gray). }
\label{tbl:lit}
\end{table*}

\section{Why Early Defect Prediction Might Work} \label{tion:whyearly}

\subsection{GitHub Results}

 Recently (2020), Shrikanth and Menzies found defect-prediction beliefs not supported by available evidence~\cite{shrikanth2019assessing}. We looked for why such confusions exist -- which lead to the discovery that pattern in \fig{whale} of project data changes dramatically over the life-cycle. \fig{whale} shows  data from 1.2m GitHub commits from 155
popular GitHub projects (the criteria for selecting those particular projects is detailed below). 
Note how the frequency defect data (shown in red/shaded) starts collapsing early in the life cycle (after 12 months). This observation suggests that it might be relatively uninformative to learn from later life cycle data. 
This was an interesting finding since, as mentioned in the introduction, 
it is common practice in defect prediction to perform ``recent validation'' where predictors are tested on the latest release after training from the prior one or two releases~\cite{tan2015online,mcintosh2017fix,kondo2020impact}. 
 In terms of 
 \fig{whale}, that strategy would train on red dots (shaded) taken near the right-hand-side, then test on the most right-hand-side dot.
 Given the shallowness of the defect data in that region, such recent validation could lead to results that are not representative of the whole life cycle.
 
 Accordingly, we sat out to determine how different training and testing sampling policies across the life cycle of 
 \fig{whale} affected the results.
 After much experimentation (described below), we assert that if data is collected up until the vertical green line of 
  \fig{whale}, then that generates a model as good as anything else.

\subsection{Related Work}\label{tion:related}

Before moving on, we first discuss related work on early life cycle defect prediction.
In 2008, Fenton et al.~\cite{fenton2008effectiveness} explored the use of human judgment (rather than data collected from the domain) to handcraft a causal model to predict residual defects (defects caught during independent testing or operational usage)~\cite{fenton2008effectiveness}. 
Fenton needed two years of expert interaction to build models that compete
with defect predictors learned by data miners from domain data.
Hence we do not explore those methods here since they were very labor-intensive.

In 2010, Zhang and Wu showed that it is possible to estimate the project quality with fewer programs sampled from an entire space of programs (covering the entire project life-cycle)~\cite{zhang2010sampling}. Although we too draw fewer samples (commits), we sample them `early' in the project life-cycle to build defect prediction models. In another 2013 study about sample size, Rahman et al. stress the importance of using a large sample size to overcome bias in defect prediction models~\cite{rahman2013sample}.
We find our proposed `data-lite' approach performs similar to `data-hungry' approaches while we do not deny bias in defect prediction data sets. Our proposed approach and recent defect prediction work handle bias by balancing defective and non-defective samples~\cite{bennin2019relative,agrawal2018better} (class-imbalance).

Recently (2020), Arokiam and Jeremy~\cite{Arokiam2020}  explored bug severity prediction. They show it is possible to predict bug-severity early in the project development by using data transferred from other projects~\cite{Arokiam2020}.
Their analysis was on the cross-projects, but unlike this paper,
they did not explore just how early in the life cycle 
did within project data became effective. In similar work to  Arokiam and Jeremy, in 2020, Sousuke ~\cite{amasaki2020cross}  explored another early life cycle, Cross-Version defect prediction (CVDP) using Cross-Project Defect Prediction (CPDP) data. Their study was not as extensive as ours (only 41 releases).
CVDP uses the project's prior releases to build defect prediction models. Sousuke compared defect prediction models trained using three within project scenarios (recent project release, all past releases, and earliest project release) to endorse recent project release. Sousuke also combined CVDP scenarios using CPDP (24 approaches) to recommend that the recent project release was still better than most CPDP approaches. However, unlike Sousuke, we offer contrary evidence in this work, as our endorsed policy based on earlier commits works similar to all other prevalent policies (including the most recent release) reported in the literature. Notably, we assess our approach on 1000+ releases and evaluate on seven performance measures.

In summary, as far as we can tell, ours is the first study to perform an extensive comparison of prevalent sampling policies practiced in the defect prediction space. 

\section{Sampling Polices}\label{tion:sp}
One way to summarize this paper is to evaluate a novel ``stop early'' sampling policy for collecting the data needed for defect prediction. 
This section describes a survey of sampling policies in defect prediction. Each sampling policy has its way of extracting training and test data from a project. As shown below, there is a remarkably diverse number of policies in the literature that have not been systematically and comparatively evaluated prior to this paper.

In April 2020, we found 737 articles in Google Scholar 
using the query (``software'' AND ``defect prediction'' AND ``just in time'', ``software'' AND `` defect prediction'' AND ``sampling policy''). 
``Just in time (JIT)'' defect prediction is a widely-used approach where the code seen in each commit
is assessed for its defect proneness~\cite{kamei2012large,fukushima2014empirical,kondo2020impact,tan2015online}.

From the results of that query, we applied some temporal filtering: (1)~we examined all articles more recent than 2017; (ii)~for older articles, we examined all papers from the last 15 years with more than
10 citations per year. After reading the title, abstracts, and the methodology sections, we found the 39 articles of \tbl{lit} that argued for particular sampling policies.

\begin{figure}[!t]
\begin{center}
\includegraphics[width=2.8in,keepaspectratio]{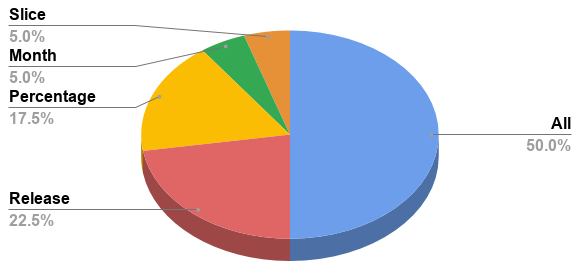}
\end{center}
\caption{Summary of sampling types from   \tbl{lit}.}\label{fig:pie}  
\end{figure}

 \fig{pie} shows a high-level view of the sampling policies seen in the  \tbl{lit} papers:
 
\begin{itemize}
 \item \textbf{All}: When the historical data (commits/files/modules etc) is used for evaluation within some cross-validation study (where the data
 is divided randomly into $N$ bins and the data from bin `$i \in N$' is used to test a model trained from all other data)~\cite{d2012evaluating}.
    \item \textbf{Percentage}: The historical data is stratified by some percentage, like 80-20\%. The minimum \% we found was 67\%~\cite{Rebalancing2018}.
    \item \textbf{Release}: The models are trained on the immediate or more past releases in order to predict defects on the current release~\cite{bennin2019relative}.
    \item \textbf{Month}: When 3 or 6 months of historical data is used to predict defects in future files, commits, or release~\cite{mcintosh2017fix}.
    \item \textbf{Slice}:  An arbitrary stratification is used to divide the data based on a specific number of days (like 180 days or six months in~\cite{zhang2014towards}). 
    
\end{itemize}

\begin{figure}[!t]
\begin{center}
  \includegraphics[width=2.9in,keepaspectratio]{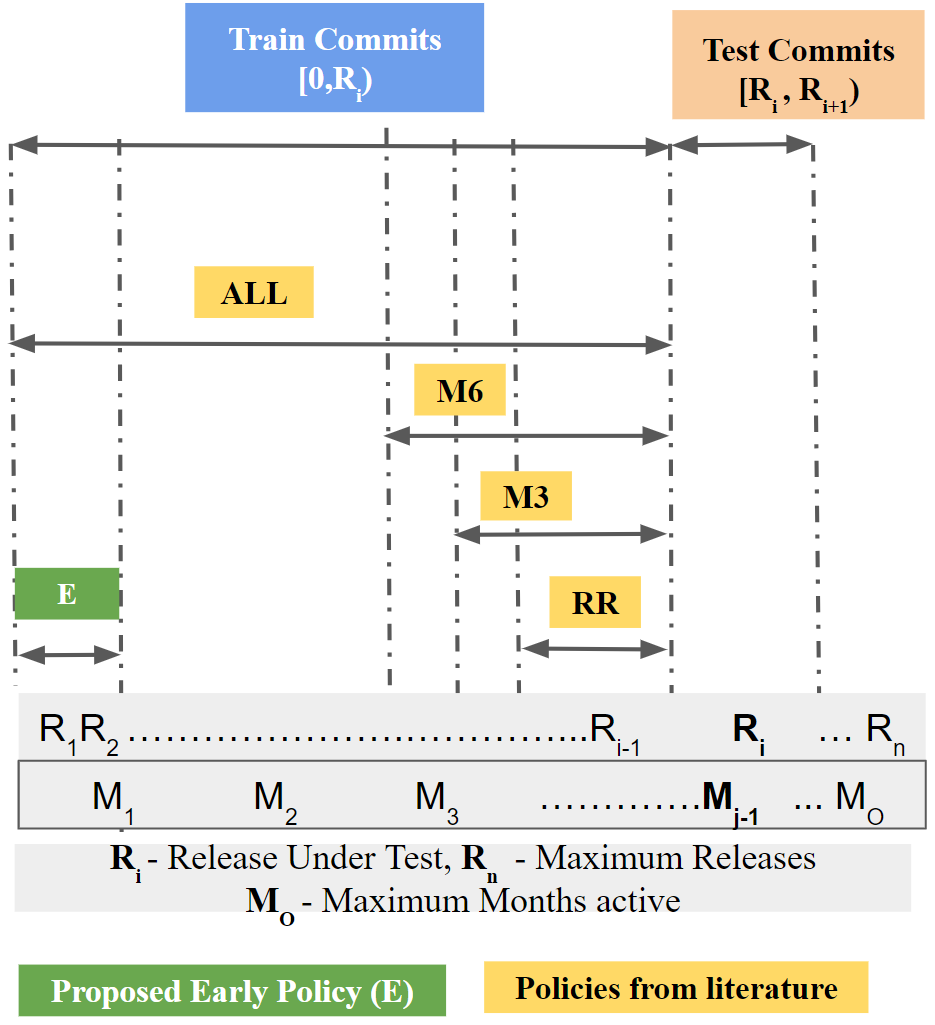}
\caption{A visual map of sampling.  Project
time-line   divided into `Train commits' and `Test  commits'. Learners learn from `Train' to classify defective commits in the `Test'.
}\label{fig:policies}  
\end{center}
\end{figure}
It turns out \fig{pie} is only an approximation of the diverse number sampling policies we see in the literature.
A more comprehensive picture is shown in \fig{policies} where we divide software releases $R_i$ that occur over many months $M_j$  into some {\em train} and {\em test} set.

  \begin{table*}[!t]
  \centering
  \caption{Four representative sampling policies from literature and an early life cycle policies (the row shown in gray).}
\begin{tabular}{|l|p{16cm}|}
\rc 
\hline
\textbf{Policy} & \textbf{Method} \\ \hline
\textbf{ALL}             & Train using all past software commits ($[0,R_{i})$) in the project before the first commit in the release under test $R_{i}$. \\
\textbf{M6}              & Train using the recent six months of software commits ($[R_{i} - 6months)$) made before the first commit in the  release under test $R_{i}$.       \\
\textbf{M3}              & Train using the recent three months of software commits ($[R_{i} - 3months)$) made before the first commit in the release under test $R_{i}$.     \\ 
\textbf{RR}              & Train  using the software commits in the previous release $R_{i-1}$ before the first commit in the  release under test $R_{i}$.                    \\ \hline
\rc \textbf{E}           & Train using early 50 commits (25 clean and 25 defective) randomly sampled within the first 150 commits before the first commit in the  release under test $R_{i}$. \\ \hline
\end{tabular}
\label{tbl:policies}
\end{table*}

 Using a little engineering judgment and guided by the frequency 
of different policies (from  \fig{pie}), we elected to focus on     four sampling policies from 
the literature and 
one `early stopping' policy,
see \tbl{policies}). The \% share in \fig{pie} show `ALL and RR' are prevalent practices whereas `M3 and M6' though not prevalent are used in related  literature~\cite{mcintosh2017fix,8816772}. We did not consider separate policies for `Percentage' and `Slice' as the former is similar to `ALL' (100\% of historical data), and the latter is least prevalent and similar to M6 (180 days or six months).

\noindent

Note the ``magic numbers'' in  \tbl{policies}: 
\bi
\item {\em 3 months, 6 months}: these are thresholds often seen in the literature.
\item {\em 25 clean + 25 defective commits}:  We arrived at these numbers based on the work of
Nam et al. built defect prediction models for using just 50 samples~\cite{nam2017heterogeneous}. 
\item
{\em Sampling at random from the first 150 commits.} Here, we did some experiments recursively dividing the data in half until defect prediction stopped working.
\ei
We will show below that early sampling  (shown in gray in \tbl{policies}) works just as well as the other policies.

\section{Methods}

\subsection{Data} \label{tion:filtering}
This section describes the data used in this study as well as what we mean by ``clean'' and ``defective'' commits.

All our data comes from open source (OS)  GitHub projects~\cite{cite_github} that we mined randomly using  \textit{Commit Guru}~\cite{rosen2015commit}. \textit{Commit Guru} is a publicly available tool based on a 2015 ESEC/FSE paper used in numerous prior works~\cite{xia2016predicting,kondo2020impact}. \textit{Commit Guru} provides a portal where it takes a request (URL) to process a GitHub repository. It extracts all commits and their features to be exported to a file. Commits are categorized (based on the occurrence of certain keywords) similar to the approach in SZZ algorithm~\cite{cite_szz_original}. The ``defective'' (bug-inducing) commits are traced using the git diff (show changes between commits) feature from bug fixing commits; the rest are labeled ``clean''.

But data from \textit{Commit Guru} does not contain release information, which we extract separately from the project tags. Then we use scripts to associate the commits to the release dates. Then those codes associated with those changes were then summarized by \textit{Commit Guru} using the attributes of \tbl{metrics}. Those attributes became the independent attributes used in our analysis. Note that the use of these particular attributes has been endorsed by prior studies ~\cite{kamei2012large,rahman2013and}. 

SE researchers have warned against using all GitHub data since this website contains many projects that might be categories
as ``non-serious'' (such as homework assignments). Accordingly, following the advice
of prior researchers~\cite{yan2020just,munaiah2017curating}, we ignored projects with
\bi
\item Less than  1000 stars;
\item Less than 1\% defects;
\item Less than two releases;
\item Less than one year of activity;
\item No license information.
\item Less than 5 defective and 5 clean commits.
\ei
 This resulted in 155 projects developed written in many languages across various domains, as discussed in \tion{introduction}.

\begin{table}
\caption{14 Commit level features that Commit Guru tool \cite{kamei2012large,rosen2015commit} mines from GitHub repositories}
\label{tbl:metrics}
\scriptsize
\begin{tabular}{|l|l|p{5cm}|}
\hline
\rowcolor[HTML]{C0C0C0} 
\textbf{Dimension}                    & \textbf{Feature} & \textbf{Definition}                                           \\ \hline
                                      & NS               & Number of modified subsystems                                 \\ \cline{2-3} 
                                      & ND               & Number of modified directories                                \\ \cline{2-3} 
                                      & NF               & Number of modified Files                                      \\ \cline{2-3} 
\multirow{-4}{*}{\textbf{Diffusion}}  & ENTROPY          & Distribution of modified code across each file                \\ \hline
                                      & LA               & Lines of code added                                           \\ \cline{2-3} 
                                      & LD               & Lines of code deleted                                         \\ \cline{2-3} 
\multirow{-3}{*}{\textbf{Size}}       & LT               & Lines of code in a file before the change                          \\ \hline
\textbf{Purpose} & FIX & Whether the change is defect Changes that fixing the defect are more likely to introduce more defects fixing ? \\ \hline
                                      & NDEV             & Number of developers that changed the modified files          \\ \cline{2-3} 
                                      & AGE              & The average time interval from the last to the current change \\ \cline{2-3} 
\multirow{-3}{*}{\textbf{History}}    & NUC              & Number of unique changes to the modified files before         \\ \hline
                                      & EXP              & Developer experience                                          \\ \cline{2-3} 
                                      & REXP             & Recent developer experience                                   \\ \cline{2-3} 
\multirow{-3}{*}{\textbf{Experience}} & SEXP             & Developer experience on a subsystem                           \\ \hline
\end{tabular}
\end{table}
\begin{figure}[!t]
\begin{center}
 \includegraphics[width=3.25in,keepaspectratio]{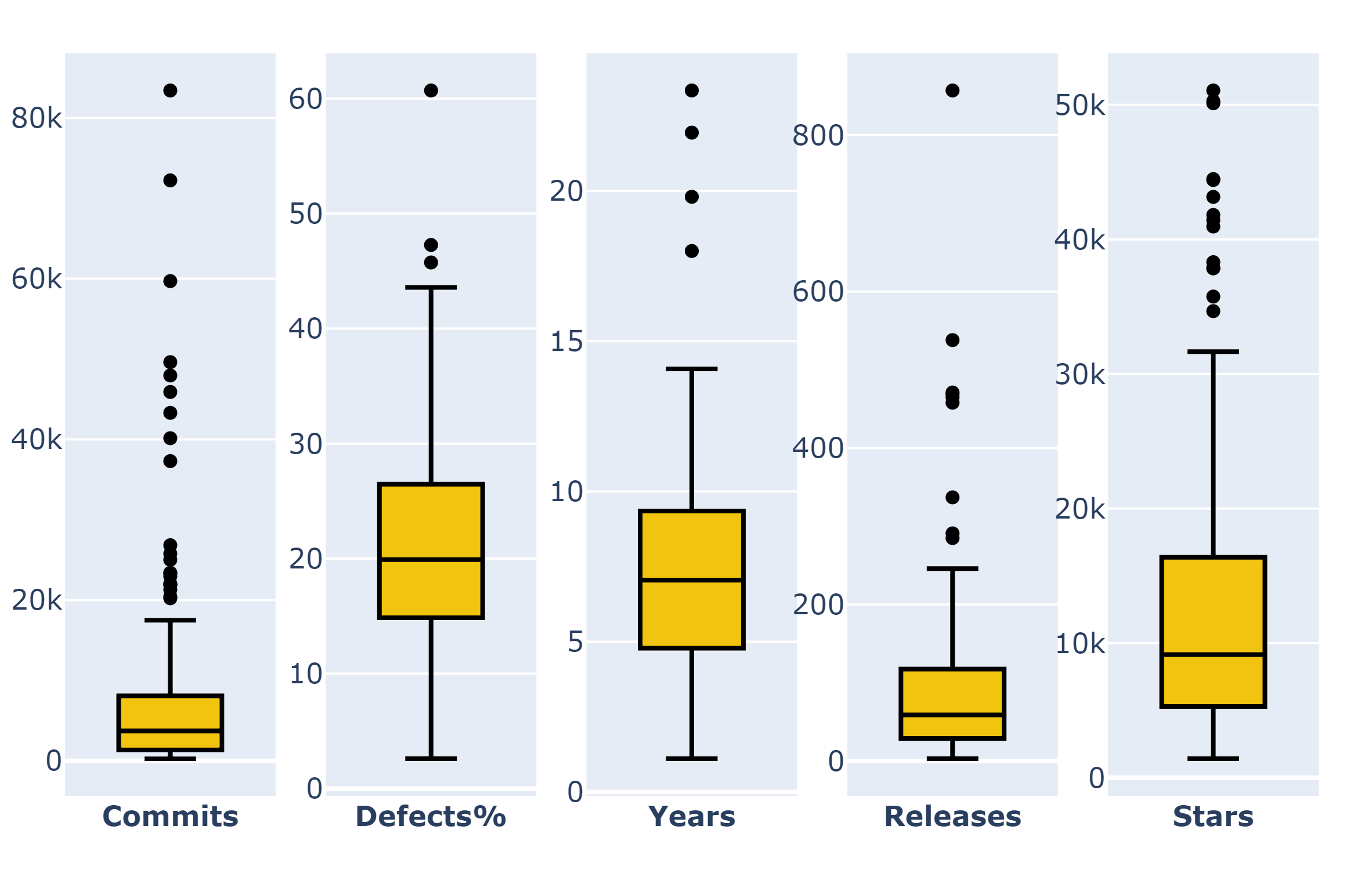}  
 \caption{Distributions seen in   all 1.2 millions commits of all 155 projects: median values of commits (3,728), percent of defective commits (20\%), 
life span in years (7), releases (59) and stars (9,149).}\label{fig:subject_systems_distribution}
\end{center}
\end{figure}
\begin{figure}[!t]
 ~~~~\includegraphics[width=2.5in,keepaspectratio]{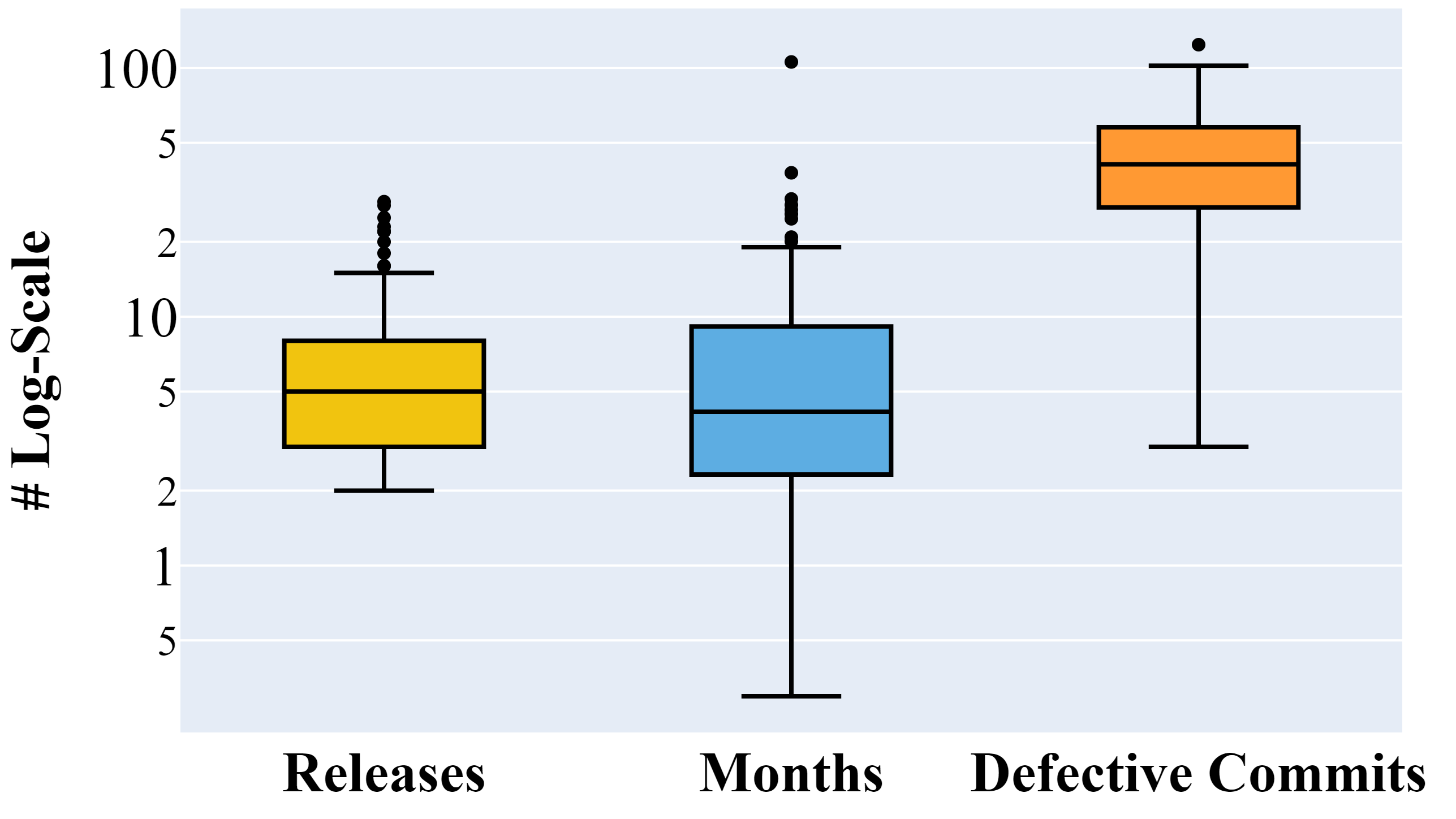}
\caption{Distributions seen in   the first 150 commits of all 155 projects;
median values  of project releases (5), project development months (4) and defective commits (41) }\label{fig:commits_timeline}  
\end{figure}

\fig{subject_systems_distribution} shows information on our selected projects. As shown in that figure, our projects have:
\bi
\item Median life spans of 84 months  with 59 releases;
\item The projects have (265, 3,728, 83,409) commits  (min, median, max) with data up to  December 2019;
\item 20\% (median) of project commits introduce bugs.  
\ei
\definecolor{ao(english)}{rgb}{0.0, 0.5, 0.0}
\definecolor{applegreen}{rgb}{0.55, 0.71, 0.0}
\setlength\fboxsep{2pt}
\fig{commits_timeline} focuses on just the data used in the early life cycle 
''\colorbox{applegreen}{\textcolor{white}{~{\bf E}~}}''
sampler described in \fig{policies}.
In the median case, by the time we can collect 150 commits, projects have had five releases in 4 months (median values).

\subsection{Algorithms}\label{tion:methodology}
 
 Our study uses three sets of algorithms:
 \bi
 \item
 The five sampling policies described above;
 \item
 The six classifiers described here;
 \item
Pre-processing for some of the sampling policies.
 \ei

\subsubsection{Classifiers}

 After an extensive analysis, Ghotra et al. ~\cite{ghotra2015revisiting} rank over 30 defect prediction algorithms into four ranks. For our work, we take six of their learners that are widely used in the literature and which can be found at all four ranks of the
Ghtora et al., study.
Those learners were:
\bi
\item Logistic Regression (LR);
\item Nearest neighbour  (KNN) (minimum 5 neighbors);
\item Decision Tree (DT);
\item  Random Forrest (RF)
\item Na\"ive Bayes (NB);
\item Support Vector Machines (SVM)
\ei

\subsubsection{Pre-processers}\label{tion:pre}
Following some advice from the literature,
we applied some  feature
engineering to the Table~\ref{tbl:metrics} data.
Based on  advice by Nagappan and Ball, we generated
relative churn and normalized LA and LD attributes by
dividing by LT and LT and NUC dividing by NF~\cite{nagappan2005use}. Also, we dropped ND and REXP since Kamei et al. reported that  NF and ND are highly correlated with REXP and EXP. Lastly, we applied the logarithmic transform to the remaining process measures (except the boolean variable `FIX') to alleviate skewness~\cite{shihab2010understanding}. 

In other pre-processing steps, we applied Correlation-based Feature Selection (CFS). Our initial experiments with this data set lead to unpromising results (recalls less than 40\%, high false alarms).
However, those results improved after we applied {\em feature subset selection} to remove spurious attributes.
CFS is a widely applied feature subset selection method proposed by Hall ~\cite{hall2003benchmarking} and is recommended in building supervised defect prediction models~\cite{kondo2019impact}. CFS is a heuristic-based method to find (evaluate) a subset of features incrementally. CFS performs a best-first search to find influential sets of features that are not correlated with each other, however, correlated with the classification. Each subset is computed as follows:
$\mathit{merit}s = kr{\mathit{cf}}/\sqrt{k+k(k-1)r_{\mathit{ff}}}$
where:
\bi
\item
$\mathit{merit}s$ is the value of  subset $s$ with $k$ features; 
\item
$r{\mathit{cf}}$ is a score that explains the connection of that feature set to the class;
\item
$r{\mathit{ff}}$ is the feature to feature mean and 
connection between the items in $s$, where $r{\mathit{cf}}$ should be large
and $r_{\mathit{ff}}$.
\ei
Another pre-processor that was applied to some sampling policies was 
{\em Synthetic Minority Over-Sampling}, or SMOTE. When the proportion of defective and clean commits (or modules, files, etc.) is not equal, learners can struggle to find the target class.
 SMOTE, proposed by Chawla et al. ~\cite{chawla2002smote} is often applied in defect prediction literature to overcome this problem~\cite{agrawal2018better,tantithamthavorn2018impact}. To achieve balance, SMOTE artificially synthesizes examples (commits) extrapolating using K-nearest neighbors (minimum five commits required) in the data set (training commits in our case)~\cite{chawla2002smote}. 
 Note that:
 \bi
 \item
 We do not apply SMOTE to policies that 
 already guarantee class balancing. For example,
 our preferred early life-cycle method selects at random 25 defective, and 25 non-defective (clean) commits from the first 150 commits. 
\item
Also, just to document that we avoided a potential methodological error~\cite{agrawal2018better}, we record here that we applied SMOTE
to the training data, but never the
test data.
\ei

\subsection{Evaluation Criteria}\label{tion:measures}

Defect prediction studies evaluated their model performance using a variety of criteria.  From the literature, we used
what we judged to be the
most  widely-used measures~\cite{
menzies2008implications,%
wang2013using,%
tantithamthavorn2018impact,%
bennin2019relative,%
zhang2014towards,%
mcintosh2017fix,%
Rebalancing2018,%
kondo2020impact,%
yatish2019mining,%
yan2019characterizing,%
7539677,%
d2012evaluating,%
kamei2012large}.
For the following seven criteria:
\bi
\item Nearly all  have the range 0 to 1 (except Initial number of False Alarms, which can be any positive number);
\item Four of these criteria need to be minimized: {\em D2H, IFA, Brier, PF}; i.e., for these criteria {\em less} is {\em better}.
\item   Three of these   criteria need to  be maximized: {\em AUC, Recall, G-Measure};
i.e. for these criteria {\em more} is {\em better}.
\ei
One reason we avoid precision is that prior work shows this measure has significant issues for unbalanced data~\cite{menzies2008implications}.

\subsubsection{Brier} Recent defect prediction papers~\cite{mcintosh2017fix,tantithamthavorn2018impact,Rebalancing2018,kondo2020impact} measure the model performance using the Brier
absolute predictive accuracy measure. Let  $C$ be the total number
of the test commits.
Let $y_{i}$ be 1 (for defective commits) or 0 otherwise.
Let  $\hat{y_{i}}$ be  the probability of commit being defective
(calculated from the loss functions in   scikit-learn classifiers~\cite{scikit-learn}).
Then:

\begin{equation}
\footnotesize
\mathit{Brier} = \frac{1}{C}\sum\limits _{t=1}^{C}(y_i-\hat{y_i})^2 \,\! \
\end{equation}

\subsubsection{Initial number of False Alarms (IFA)} Parnin and Orso ~\cite{parnin2011automated}  say that developers lose faith in analytics if they see too many initial false alarms.  IFA is simply the number of false alarms encountered after sorting the commits in the order of probability of being detective, then counting the number of false alarms before finding the first true alarm.

\subsubsection{Recall} Recall is the proportion of inspected defective commits among all the actual defective commits. 

\begin{equation}
\footnotesize
\mathit{Recall} = \frac{\mathit{True\ Positives}}{\mathit{True\ Positives + False\ Negatives}} 
\end{equation}

\subsubsection{False Positive Rate (PF)} 
The proportion of predicted defective
commits those are not defective among all
the predicted defective commits.

\begin{equation}
\footnotesize
\mathit{PF} = \frac{\mathit{False\ Positives}}{\mathit{False\ Positives + True\ Negatives}}
\end{equation}

\subsubsection{Area Under the Receiver Operating Characteristic curve (AUC)} AUC is the area under the curve between the true positive rate and false-positive rate.

\subsubsection{Distance to Heaven (D2H)} D2H or ``distance to heaven'' aggregates on two metrics Recall and False Positive Rate (PF) to show how close to  ``heaven'' (Recall=1 and PF=0)~\cite{chen2018applications}.  

\begin{equation}
\footnotesize
\mathit{D2H}  = \frac{\sqrt{(1-\mathit{Recall})^2 + (0-\mathit{PF   })^2}}{ \sqrt{2}}
\end{equation}

\subsubsection{G-measure (GM)} 
A harmonic mean between Recall and the compliment of PF measured, as shown below.
\begin{equation}
\footnotesize
\mathit{G-Measure}  = \frac{2 * \mathit{Recall} * (1-\mathit{PF})}{ \mathit{Recall}+(1-\mathit{PF})}
\end{equation}
Even though GM and D2H combined the same underlying measures,
we include both here since they both have been used separately in the literature. Also, as shown below, it is not necessarily true
that achieving good results on GM   means that good results will also be achieved with D2H.

Due to the nature of the classification process,
 some criteria will  always offer contradictory results:
 \bi
 \item
 A learner can achieve 100\% recall just by declaring that all examples belong to the target class.
 This method
 will incur a high false alarm rate.
\item
 A learner can achieve 0\% false alarms just by declaring that no examples belong to the target class.
 This method
 will incur a very low recall rate.
 \item
 Similarly, Brier and Recall
 are also antithetical since
 reducing the loss function also
 means missing some conclusions
 and lowering recall.
 \ei
 
\subsection{Statistical Test}\label{tion:sk}


Later in \tion{results}, we compare distributions of evaluation measures of various sampling policies that may have the same median while their distribution could be very different. Hence to identify significant differences (rank) among two or more populations, we use the Scott-Knott test recommended by Mittas et al. in TSE'13 paper~\cite{Mittas13}. This test is a top-down bi-clustering approach for ranking different treatments, sampling policies in our case. This method sorts a list of $l$ sampling policy evaluations with $\mathit{ls}$ measurements by their median score. It then splits $l$ into sub-lists \textit{m, n} in order to maximize the expected value of differences in the observed performances before and after divisions. 

For lists $l,m,n$ of size $\mathit{ls},\mathit{ms},\mathit{ns}$ where $l=m\cup n$, the ``best'' division maximizes $E (\Delta)$; i.e.
the difference in the expected mean value
before and after the spit: 
\[E (\Delta)=\frac{ms}{ls}abs (m.\mu - l.\mu)^2 + \frac{ns}{ls}abs (n.\mu - l.\mu)^2\]

We also employ the conjunction of bootstrapping and A12 effect size test by Vargha and Delaney~\cite{vargha2000critique} to avoid ``small effects'' with statistically significant results.  

Important note: we apply our statistical methods
separately to all the evaluation criteria; i.e., when we compute ranks, we do so for (say) false alarms separately to recall.  

\subsection{Experimental Rig}\label{tion:rig}
By definition, our different sampling policies have 
{\em different} train and {\em different} test sets.  But, methodologically, when we compare these different policies, we have to compare results on the {\em same} sets of releases. 
To handle this we:
\bi
\item
First, run all our six policies, combined with all our six learners. This offers multiple predictions to different commits.
\item
Next, for each release, we divide the predictions into those that come from the same learner:policy pair. These divisions are then assessed with statistical methods described above.
\ei

\begin{table*}[!b]
\setlength\extrarowheight{2pt} 
 \begin{center}
 \scriptsize
\begin{tabular}{|l|l|l|rrrrrrr|}  \hline 
\rc \textbf{Policy} & \textbf{Classifier} & \textbf{Wins} & \textbf{D2H-} & \textbf{AUC+} & \textbf{IFA-} & \textbf{Brier-} & \textbf{Recall+} & \textbf{PF-} & \textbf{GM+} \\ \hline 
 
\all M6           & NB                  & \multirow{5}{*}{4} & \hil 0.37    & \hil 0.67    & 1.0          & 0.32           & \hil 0.78       & 0.33        & \hil 0.67        \\
\all M3           & NB                  &                    & \hil 0.37    & \hil 0.67    & 1.0          & 0.31           & \hil 0.76       & 0.32        & \hil 0.68        \\
\early E & LR                  &                    & \hil 0.36    & \hil 0.68    & 1.0          & 0.32           & \hil 0.71       & 0.31        & \hil 0.68        \\
\all M6           & SVM                 &                    & \hil 0.43    & 0.65         & \hil 0.0     & \hil 0.21      & 0.44            & \hil 0.1    & 0.48             \\
\all M3           & SVM                 &                    & \hil 0.43    & 0.65         & \hil 0.0     & \hil 0.21      & 0.43            & \hil 0.1    & 0.48             \\ \hline
\all ALL         & NB                  & 3                  & \hil 0.4     & 0.65         & 1.0          & 0.36           & \hil 0.83       & 0.40         & \hil 0.67        \\ \hline
\early E & KNN                 & \multirow{8}{*}{2} & \hil 0.39    & 0.65         & 1.0          & 0.33           & 0.65            & 0.32        & \hil 0.62        \\
\all ALL         & LR                  &                    & \hil 0.38    & 0.66         & 1.0          & 0.3            & 0.65            & 0.25        & \hil 0.62        \\
\all M6           & LR                  &                    & \hil 0.36    & \hil 0.68    & 1.0          & 0.25           & 0.59            & 0.19        & 0.60              \\
\all M3           & LR                  &                    & \hil 0.36    & \hil 0.68    & 1.0          & 0.24           & 0.58            & 0.17        & 0.60              \\
\all M6           & KNN                 &                    & \hil 0.4     & 0.65         & \hil 0.0     & 0.23           & 0.50             & 0.14        & 0.53             \\
\all ALL         & SVM                 &                    & \hil 0.4     & 0.66         & \hil 0.0     & 0.25           & 0.50             & 0.14        & 0.54             \\
\all M3           & KNN                 &                    & \hil 0.41    & 0.65         & \hil 0.0     & 0.23           & 0.47            & 0.13        & 0.51             \\
\all M6           & RF                  &                    & 0.44         & 0.63         & \hil 0.0     & 0.24           & 0.43            & \hil 0.12   & 0.47             \\ \hline
\early E & SVM                 & \multirow{7}{*}{1} & \hil 0.4     & 0.64         & 1.0          & 0.31           & 0.6             & 0.26        & 0.59             \\
\all ALL         & KNN                 &                    & \hil 0.38    & 0.66         & 1.0          & 0.25           & 0.55            & 0.17        & 0.57             \\
\all ALL         & DT                  &                    & \hil 0.42    & 0.62         & 1.0          & 0.32           & 0.52            & 0.25        & 0.54             \\
\all M6           & DT                  &                    & \hil 0.43    & 0.62         & 1.0          & 0.29           & 0.5             & 0.2         & 0.51             \\
\all ALL         & RF                  &                    & \hil 0.42    & 0.64         & 1.0          & 0.26           & 0.49            & 0.15        & 0.51             \\
\all M3           & DT                  &                    & \hil 0.43    & 0.62         & 1.0          & 0.28           & 0.48            & 0.19        & 0.5              \\
\all M3           & RF                  &                    & 0.44         & 0.63         & 1.0          & 0.24           & 0.42            & \hil 0.11   & 0.46             \\  \hline
\early E & DT                  & \multirow{3}{*}{0} & 0.46         & 0.58         & 1.0          & 0.38           & 0.57            & 0.35        & 0.54             \\
\early E & NB                  &                    & 0.54         & 0.54         & 1.0          & 0.37           & 0.55            & 0.29        & 0.41             \\
\early E & RF                  &                    & 0.44         & 0.61         & 1.0          & 0.33           & 0.52            & 0.26        & 0.52                \\  \hline
 
\end{tabular}
 \setlength\extrarowheight{0pt}
~\\~\\KEY: \ \fcolorbox{all}{all}{} More data (ALL,M6 and M3) 
\fcolorbox{early}{early}{} Early (E)
\end{center}
\caption{24 defect prediction models tested in all 4,876 applicable project releases.  In the first row ``+'' and ``-'' denote
 the criteria that need to be maximized or minimized, respectively.
`Wins' is the frequency of the policy found in the top \#1 Scott-Knott rank in each of the seven evaluation measures  (the cells shaded in gray).  
}
\label{tbl:more}
\end{table*}
\begin{table*}[!b]

 \begin{center}
\setlength\extrarowheight{2pt} 
\scriptsize
\begin{tabular}{|l|l|l|rrrrrrr|}  
\hline
\rc \textbf{Policy}        & \textbf{Classifier} & \textbf{Wins} & \textbf{D2H-} & \textbf{AUC+} & \textbf{IFA-} & \textbf{Brier-} & \textbf{Recall+} & \textbf{PF-} & \textbf{GM+} \\ \hline

\early E & LR         & 4    & \hil 0.36 & \hil 0.68 & 1.0      & 0.32      & \hil 0.71 & 0.31      & \hil 0.68 \\ \hline
\rr RR           & NB         & 3    & \hil 0.38 & \hil 0.66 & 1.0      & 0.32      & \hil 0.71 & 0.30       & 0.65      \\
\rr RR           & LR         & 3    & \hil 0.35 & \hil 0.68 & 1.0      & \hil 0.24 & 0.59      & 0.18      & 0.61      \\
\rr RR           & SVM        & 3    & 0.42      & 0.64      & \hil 0.0 & \hil 0.23 & 0.47      & \hil 0.12 & 0.5       \\ \hline
\early E & KNN        & 2    & \hil 0.39 & 0.64      & 1.0      & 0.34      & \hil 0.64 & 0.32      & 0.62      \\
\rr RR           & KNN        & 2    & 0.41      & 0.64      & 1.0      & \hil 0.25 & 0.5       & \hil 0.15 & 0.53      \\
\rr RR           & RF         & 2    & 0.43      & 0.63      & 1.0      & \hil 0.24 & 0.43      & \hil 0.13 & 0.48      \\
\early E & SVM        & 1    & \hil 0.4  & 0.64      & 1.0      & 0.31      & 0.6       & 0.26      & 0.59      \\ \hline
\early E & DT         & 0    & 0.46      & 0.58      & 1.0      & 0.39      & 0.56      & 0.35      & 0.54      \\
\early E & NB         & 0    & 0.54      & 0.54      & 1.0      & 0.37      & 0.54      & 0.29      & 0.42      \\
\early E & RF         & 0    & 0.44      & 0.61      & 1.0      & 0.33      & 0.52      & 0.26      & 0.53      \\
\rr RR           & DT         & 0    & 0.42      & 0.62      & 1.0      & 0.28      & 0.50       & 0.20       & 0.51     \\ \hline
\end{tabular}
~\\~\\KEY:  \fcolorbox{recency}{recency}{} Recency (RR)  \fcolorbox{early}{early}{} Early (E)
\end{center}
\caption{12 defect prediction models tested on 3,704 project releases.
 In the first row ``+'' and ``-'' denote
criteria that need to be maximized or minimized, respectively.
`Wins' is the frequency of the policy found in the top \#1 Scott-Knott rank in each of the seven evaluation measures  (the cells shaded in gray). 
 }
\label{tbl:recency}
\end{table*}

\section{Results}\label{tion:results}

Tables~\ref{tbl:more} and \ref{tbl:recency}  show results
when our six learners applied our five sampling policies. 
We splot these results into two tables since
Our policies lead to results with   different samples sizes:
the recent release, or ``RR'',
the policy uses data from just two releases while ``ALL''
uses everything.

 In the first row of those tables, ``+'' and ``-'' denote
criteria that need to be maximized or minimized, respectively.
Within  the tables, \colorbox{gray!25}{gray cells} show
statistical test results (conducted separately on each criterion).  Anything
ranked ``best'' is colored gray, while all else 
have white backgrounds.

Columns one and two show the policy/learners that lead to these results. Rows are sorted by how often policy/learners ``win''; i.e., achieve
best ranks across all criteria.
In 
Tables~\ref{tbl:more} and \ref{tbl:recency}, no policy+learner wins 7 out of 7 times on all criteria,
so some judgment will be required
to make an overall conclusion.
Specifically,
based on results
from the multi-objective
optimization literature,
we will first remove the policies+learners that score
worse on most criteria, then debate
trade-offs between the rest.

To simplify that trade-off debate, we offer two notes.
Firstly, cross all our learners, the median value for IFA is very
small-- zero or one; i.e., developers using
these tools only need to suffer one false alarm or less before finding something they need to fix.
Since these observed IFA scores are so small, we say that ``losing'' on IFA is hardly a reason to dismiss a learner/sampler combination.
Secondly,
 D2H and GM combine multiple criteria.
For example,   ``winning'' on D2H and
GM means
performing well on both Recall and PF; i.e. these
two criteria are somewhat more informative than the others.
 
Turning now to those results,
we explore  two issues.
For defect prediction:

\noindent {\bf RQ1:} Is more data, better?


\noindent {\bf RQ2:} When is more recent data better than older data?

Note that we do explore a third research issue: are different learners better at learning from a little, a lot,  or all the available data. Based on our results, we have nothing definitive to offer on that issue. That said,
if we were pressed to recommend a particular learning algorithm, then we say there are no counterexamples to the claim that  ``it is useful to apply
CFS+LR''.

\subsection*{RQ1:    Is   more   data,   better? }\label{tion:rq1}

{\em Belief1:} 
Our introduction included examples where proponents  of data-hungry methods 
advocated that if data is useful,  then even more data is much more useful.

{\em Prediction:1}
If that belief was  the case, then  in
Table~\ref{tbl:more}, 
data-hungry sampling
policies that used more
data should defeat ``data-lite'' sampling policies.

{\em Observation1a:}
In Table~\ref{tbl:more}, 
Our ``data hungriest''  sampling policy (ALL) 
loses on on most criteria.
While it achieves the highest Recall (83\%), it also has the highest false
alarm range (40\%). 
 As to  which other policy is preferred  in the  best {\em wins=4} zone of  Table~\ref{tbl:more}, there is no clear winner. What we would say here is that our preferred ``data-lite'' method 
 called ``E'' 
 (that uses 25 defective and 25 non-defective commits selected at random from the first 150 commits) is competitive with the rest.  Hence:

\begin{tcolorbox}[colback=red!5,colframe=gray!50]
\textbf{Answer1a:} For defect prediction, it is not clear that more data is inherently better.
\end{tcolorbox}

{\em Observations1b:} 
Figure~\ref{fig:subject_systems_distribution} of this paper showed that within our sample of projects, we have data lasting a median of 84 months. 
Figure~\ref{fig:commits_timeline}   noted that by the time we get to 150 commits, most projects are 4 months old (median values).
The ``E'' results of Table~\ref{tbl:more} showed that defect models learned from that 4 months of data are competitive with all the other policies studied here. Hence we say,

\begin{tcolorbox}[colback=red!5,colframe=gray!50]
\textbf{Answer1b:} 
96\% of the time,  we do not  want and
we do not  need data-hungry methods
\end{tcolorbox}

\subsection*{RQ2: When is more recent data better than older data? }\label{tion:rq3}

{\em Belief2:} As discussed earlier in our introduction, many researchers prefer using recent data over data from earlier periods. 
For example, 
it is common practice in defect prediction
 to perform ``recent  validation'' where predictors are tested on
 the latest release after training from the prior one or two
 releases~\cite{tan2015online,mcintosh2017fix,kondo2020impact,fu2016tuning}. 
For a project with multiple releases,  recent validation ignores all the insights that are available from older releases.
 
 {\em Prediction2:} If recent  data is comparatively more informative than older data, then defect predictors built
 on recent data should out-perform predictors built on much older data.

{\em Observations2:} We observe that:
\bi
\item
Figure~\ref{fig:subject_systems_distribution} of this paper showed that within our sample of projects, we have data lasting a median of 84 months.
\item
Figure~\ref{fig:commits_timeline}   noted that by the time we get to 150 commits, most projects are 4 months old (median values).
\item
 Table~\ref{tbl:recency} says that
  ``E'' wins over ``RR'' since it falls in the best
{\em wins=4} section.
\item Hence we could conclude that older data is {\em more}
effective than recent data.
\ei
That said, we feel somewhat more the circumspect conclusion is in order.
When we   compare  E+LR to the next learner in that table (RR+NB)
we only find a minimal difference in their performance scores. Hence we 
make a somewhat humbler conclusion:

\begin{tcolorbox}[colback=red!5,colframe=gray!50]
\textbf{Answer2:} Recency based methods perform no better than results from early life cycle defect predictors.
\end{tcolorbox}

This is a startling result for two reasons. Firstly, 
compared to the ``RR'' training data, the ``E'' training data is very old indeed. For projects lasting 84 months long,
``RR'' is trained on information from recent few months, with  ``E'' data comes from years before that.
Secondly,
this result calls into question any conclusion made in a paper that used recent validation to assess their approach; e.g.~\cite{tan2015online,mcintosh2017fix,kondo2020impact,fu2016tuning}.

\section{Threats to Validity}\label{tion:threats}

\subsection{Sampling Bias} \label{tion:sampling_bias}
The conclusion's generalizability will depend upon the samples considered; i.e., what matters here may not be true everywhere. To improve our conclusion's generalizability, we mined 155 long-running OS projects that are developed for disparate domains and written in numerous programming languages. Sampling trivial projects (like homework assignments) is a potential threat to our analysis. To mitigate that, we adhered to the advice from prior researchers as discussed earlier in \tion{introduction} and \tion{filtering}. We find our sample of projects have 20\% (median) defects as shown in \fig{subject_systems_distribution} nearly the same as data used by Tantithamthavorn et al.~\cite{Rebalancing2018} who report  30\%  (median) defects.

\subsection{Learner bias} Any single study can only explore a handful of classification algorithms. For building the defect predictors in this work, we elected six learners (Logistic Regression, Nearest neighbor, Decision Tree, Support Vector Machines, Random Forrest, and Na\"ive Bayes). These six learners represent a plethora of classification algorithms~\cite{ghotra2015revisiting}.

\subsection{Evaluation bias} 
We use seven evaluation measures (Recall, PF, IFA, Brier, GM, D2H, and AUC). Other prevalent measures in this defect prediction space include precision. However, as mentioned earlier, precision has issues with unbalanced data~\cite{menzies2008implications}.

\subsection{Input Bias} Our proposed sampling policy `E' randomly samples 50 commits from early 150 commits. Thus it may be true that different executions could yield different results. However, this is not a threat because each time, the early policy `E ' randomly samples 50 commits from early 150 commits to test sizeable 8,490 releases (from \tbl{more} and \tbl{recency}) across all the six learners. In other words, our conclusions about `E' hold on a large sample size of numerous releases.

\section{Conclusion} \label{tion:conclusion}

 When data keep changing, the models we can learn from that data may also change.
 If  conclusions become too fluid
 (i.e., change too often),
 then no one has a stable basis for making decisions or communicating insights.

Issues with conclusion instability disappear if,
early in the life cycle, we can learn a   predictive model that is effective for the rest of the project. 
This paper has proposed a methodology for assessing such early life cycle predictors.
\begin{enumerate}
\item Define a project selection  {\em criteria}. For this paper, our selection criteria are taken
from related work
 (from recent EMSE, TSE papers~\cite{yan2020just,munaiah2017curating});
\item Select some software analytics {\em task}. For this paper, we have explored learning defect predictors.
\item
See how early projects selected by the {\em criteria}   can be modeled for that {\em task}.
Here we found that defect predictors learned from the first four months of data perform as well as anything else.
\item
Conclude that projects matching {\em criteria} need more data for {\em task} before time found in step 3.  In this paper, we found that for 96\% of the time,  we neither want nor need data-hungry defect prediction. 
\end{enumerate} 
We stress that this result has only been shown here for defect prediction and only for the data selected by our criteria.

As for future work, we have many suggestions:
\bi
\item
The clear next step in this work is to check the validity of this conclusion
beyond the specific  {\em criteria} and {\em task} explored here.
\item
We need to revisit
all prior results that  used recent validation to assess their approach; e.g.~\cite{tan2015online,mcintosh2017fix,kondo2020impact,fu2016tuning} since our {\bf RQ2} suggests they may have been
working in a relatively uninformative region of the data.

\item
While the performance scores of Tables~\ref{tbl:more} and \ref{tbl:recency} are reasonable, there is still much room for improvement. Perhaps if we augmented early life cycle defect predictors
with a little transfer learning (from other projects~\cite{nam2017heterogeneous}), then we could generate better
performing predictors.
\item
Further to the last point,
 another interesting avenue of future research might be
hyper-parameter optimization (HPO)~\cite{tantithamthavorn2016automated,fu2016tuning,agrawal2019dodge}. HPO is often not applied in software analytics due to its computational complexity. Perhaps that complexity can be avoided by focusing only on small samples of data from very early in the life cycle.
\ei

\section*{Acknowledgements}
\noindent This work was partially supported by 
  NSF grant  \#1908762.

\bibliographystyle{IEEEtran}
 \IEEEtriggeratref{72}
\bibliography{references}

\end{document}